\documentclass{article}
\usepackage{graphicx}

\usepackage{amsmath,amssymb}
\usepackage{fontenc, times, mathptmx}
\usepackage{mathtext}
\usepackage{amstext}
\usepackage{relsize}

\usepackage{latexsym,graphicx,epsfig}
\def\bra{$\begin{array}}
 \def\era{\end{array}$}

\newcommand{\be}{\begin{equation}}
\newcommand{\ee}{\end{equation}}
\newcommand{\bea}{\begin{eqnarray}}
\newcommand{\eea}{\end{eqnarray}}
\newcommand{\bes}{\begin{subequations}}
\newcommand{\ees}{\end{subequations}}
\newcommand{\bear}{\begin{equation}\begin{array}}
\newcommand{\eear}[1]{\end{array}\label{#1}\end{equation}}
\newcommand{\fr}[2]{\dfrac{{ #1}}{{ #2}}}

\newcommand{\la}{\langle}
\newcommand{\ra}{\rangle}
\def\ba{$$\begin{array}}
\def\ea{\end{array}$$}

\renewcommand{\ge}{\geqslant}

\usepackage{bm}

\usepackage[dvips]{color}

\begin{document}

\title{ Light charged Higgs at LHC}
\author{I. F. Ginzburg\\
 Sobolev Institute of Mathematics and Novosibirsk State University\\
{\it Novosibirsk, Russia}}
\date{}

\maketitle

%\vspace{-1cm}

%%%%%%%%%%%%%%%%%%%%%%%%%%%%%%%%%%%%%%%%%%%%%%%%%%%%%%%%%%%%%%%%%%%%%
\begin{abstract}
The LHC program for discovery of light charged Higgs with mass 135-180
GeV  has to  take into account  decay $H^+\to \bar{b}\,t^*\to
W^+b\bar{b}$. The distribution of decay products in effective mass of $t^*=Wb$ is obtained.
\end{abstract}

%\pacs  _12.60Fr, 14.80 Cp

The discovery of a charged Higgs boson $H^\pm$  will be a
clear indication of non-minimal realization of Higgs mechanism.
Below we consider the Two-Higgs-Doublet Model
(2HDM) with two basic scalar doublets $\phi_1$ and $\phi_2$, in which vacuum state is characterized by two v.e.v.'s  $\la\phi_1\ra=v\cos\beta/\sqrt{2}$ and $\la\phi_2\ra=v\sin\beta/\sqrt{2}$ with $v=246$~GeV.  The observable particles  in this model are two neutral scalar Higgs
bosons $h$ and $H$ with $M_H\geqslant M_h$, pseudoscalar
$A$ and two charged Higgses $H^\pm$ with mass $M_+$. We assume, in accordance with recent LHC data, that $M_h\approx 125$~GeV.

The interaction of charged Higgs   with fermions of one generation can be parameterized by Lagrangian
 \bear{c}
\!\!\!\!L_{c}
\!=\!\fr{\sqrt{2}\left[\bar{u}\left(M_dP_R
F_D\!+\! M_uP_L F_U\right) d\!+\!\bar{\nu}M_\ell
P_R F_\ell\ell\right]H^+ \!}{v}\!+\! \!h.c.\!
 \eear{Yukser}

The 2HDM allows different variants of  the  Yukawa interaction. The most popular models for this interaction and coefficients $ F_{U,D}$ for them are summarized in the Table~\ref{Yuktype}.
The Yukawa sector in 2HDM-II is the same as in MSSM, here right down-type
\begin{table}[hbt]
\caption{\it  The most popular models of the Yukawa interactions in the 2HDM
and Yukawa couplings ${ F}_A$. The symbols u, d, $\ell$ refer to right up- and down-type quarks
and charged leptons of any generation.  }
\begin{center}\begin{tabular}{|c||c|c||c|c||c|c||}\hline
Fermion &\multicolumn{2}{|c||}{d} &\multicolumn{2}{|c||}{u} &\multicolumn{2}{|c||}{$\ell$}\\ \hline
Model&&${ F}_D$&&${ F}_U$&&${ F}_\ell$\\\hline
I &$\varphi_2$&$-\cot\beta$ &$\varphi_2$&$\cot\beta$ &$\varphi_2$&$-\cot\beta$\\ \hline
II &$\varphi_1$&$\tan\beta$ &$\varphi_2$&$-\cot\beta$ &$\varphi_1$&$\tan\beta$\\ \hline
%III&$\varphi_1\&\varphi_2$&mod.-dep. &$\varphi_1\&\varphi_2$&mod.-dep.&$\varphi_1\&\varphi_2$&mod.-dep.\\ \hline
X&$\varphi_2$&$-\cot\beta$&$\varphi_2$&$\cot\beta$&$\varphi_1$&$\tan\beta$\\ \hline
Y&$\varphi_1$&$\tan\beta$&$\varphi_2$&$\cot\beta$&$\varphi_2$&$-\cot\beta$\\ \hline
\end{tabular}\end{center}\label{Yuktype}
\end{table}
quarks and charged leptons are coupled to $\phi_1$, while right up-type quarks  are coupled to $\phi_2$. In the 2HDM-I all right quarks and leptons are coupled to $\phi_2$ only. In the 2HDM-X ({\it lepton-specific model}) all right quarks  are coupled to $\phi_2$ while right leptons are coupled to $\phi_1$ (for quark sector 2HDM-X coincides  with 2HDM-I).
In the 2HDM-Y ({\it flipped model})  right down-type quarks  are coupled to $\phi_1$, while right up-type quarks and leptons  are coupled to $\phi_2$ (for quark sector 2HDM-Y coincides with 2HDM-II). The table does not include  2HDM-III. This title combines  models in which  individual fermions interact with both scalars.

In the 2HDM-II the well-measured $B\to X_s\gamma$ decay rate excludes $M_+<300$~GeV \cite{300GEV}. For
other models $H^\pm$ can be lighter, up to 90 GeV (limitation
from LEP).

The strategy of hunting for charged Higgs depends on its mass. At $M_+>180$~GeV main decay channels of $H^+$ are $H^+\to t\bar{b}$ and $H^+\to hW^+$. At $M_+<180$~GeV these decay channels are kinematically forbidden and for the first glance main decay channels are $H^+\to c\bar{s}$, $H^+\to \tau^+\nu$.
This statement is, generally speaking, incorrect  since the coupling constant $g(H^+t\bar{b})$  is  $\sim 135$ times  larger than $g(H^+c\bar{s})$. Because of this huge strengthening, the probabilities of three-body decay  $H^+\to t^*\bar{b}\to Wb\bar{b}$ and two-particle decay $H^+\to c\bar{s}$ are comparable  (here $t^*=W+b$
is off-shell $t$).

This fact was noticed while studying  small $\tan\beta$ region of MSSM (where Higgs sector is a special case of 2HDM-II)  \cite{Djouadi:1995gv}. Later the same phenomenon was found in the 2HDM with another forms of Yukawa interaction (see e.g. \cite{ModelIXY}). Unfortunately, while discussing the discovery of the light charged Higgs at the LHC some modern papers  do not pay attention to this fact  \cite{H+conf}, \cite{Rui}.

All mentioned calculations were performed using a package  HDECAY \cite{HDECAY}. We find useful to present simple analytical equations, which allow to perform  rapid analysis with reasonable accuracy. Besides, we demonstrate that the effect takes place even at large $\tan\beta$.

First of all, for 2-body   decays we have
\bear{c}
\Gamma(H^+\to c\bar{s})=3M_+\,\fr{M_s^2 F_D^2+M_c^2 F_U^2}{8\pi v^2}\,,\quad
\Gamma(H^+\to \tau^+\nu)=M_+\fr{M_\tau^2 F_\ell^2}{8\pi v^2}\,.\eear{cs+dec}

We  denote the effective mass of  $t^*=W+b$ by $M^*$  and
\bear{c}
y=M_+^2/M_W^2,\quad  x=M^{*2}/M_W^2 \quad \mbox{with}\quad y>x>1\, .
\eear{variables}

In the calculation  of the width  $\Gamma(H^+\to t^*\bar{b}\to Wb\bar{b})$ we neglect $M_b$ in the kinematical relations and traces. The distribution in effective mass $M^*$  has form
\bear{c}
\fr{d\Gamma(H^+\to\bar{b}t^*\to \bar{b}bW^+)}{dx}
=M_+\fr{3\alpha}{(8\pi)^2s_W^2}\,\fr{M_t^2}{v^2}\times\\[3mm]\fr{(y-x)^2(x-1)^2(2+x)}{y^2x(T-x)^2}%\times\\[3mm]
\left(F_U^2+F_D^2\fr{\gamma}{x}+4F_UF_D\fr{\gamma}{y-x}\right);\\[2mm]
T\!=\!\fr{M_t^2}{M_W^2}\!=\!4.58,\;\; \gamma\!=\!\fr{M_b^2}{M_W^2}\!=\!2.7\cdot 10^{-3},\;\; s_W\!=\!\sin\theta_W.
\eear{}
All parameters were taken from ref.~\cite{PDG}. The range of parameters, describing light charged Higgs $M_+<M_t$ is   $y<T$.

Let us consider   hadronic widths in different cases.

\begin{figure}[h!]
\begin{center}
\includegraphics[height=4cm,width=0.45\textwidth]{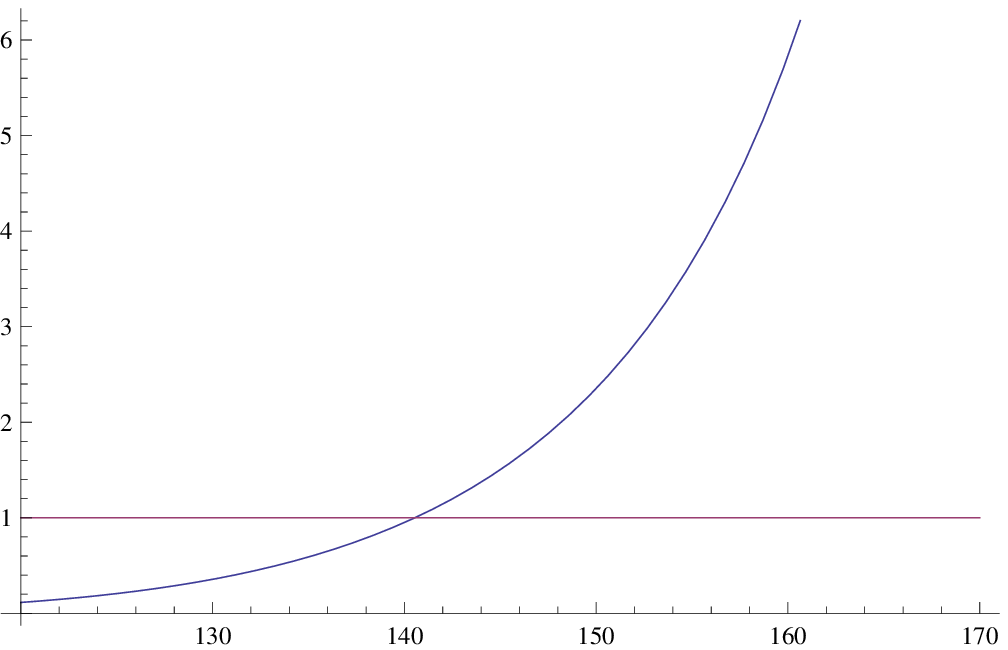}\hspace{4mm}
 \includegraphics[height=3cm,width=0.45\textwidth]{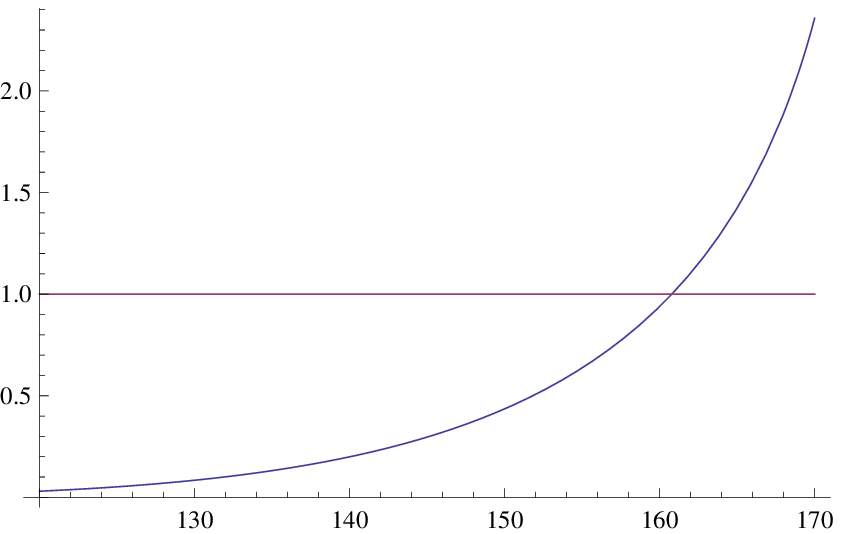}
    \caption{Ratio  $\Gamma(H^+\to t^*\bar{b}\to W^+b\bar{b})/\Gamma(H^+\to c\bar{s})$
     vs. mass $M_+$.\newline \ \   Left -- 2HDM-I, 2HDM-X and 2HDM-Y at $\tan\beta<6$.\ \  Right --  2HDM-Y at $\tan\beta>9$.}
    \label{FigratIX}
\end{center}
\end{figure}

{\bf For  2HDM-I and 2HDM-X} the coupling $F_U$ dominates at
all $\tan\beta$. Hence,  the ratio of hadronic widths is independent
on $\tan\beta$. The ratio of partial widths of $H^+$ decay is
 \bear{c} \fr{\Gamma(H^+\to\bar{b}t^*\to
\bar{b}bW^+)}{\Gamma(H^+\to
cs)}%=\\[2mm]
=\left(\fr{M_t}{M_c}\right)^2\fr{\alpha}{8\pi
s_W^2}\mathlarger\int\limits_1^y
\fr{(y-x)^2(x-1)^2(2+x)}{y^2x(T-x)^2}\,dx=\\[3mm]
=23.05\mathlarger\int\limits_1^y
\fr{(y-x)^2(x-1)^2(2+x)}{y^2x(T-x)^2}\,dx. \eear{RatioIX}
This ratio is shown in Fig.~\ref{FigratIX}-A.  One can see  that at $M_+>130$~GeV  the discussed channel reduces strongly BR's for channels $H^+\to c\bar{s}$ and $H^+\to \tau^+\nu$, at $M_+>145$~GeV the channel with off-shell $t$ becomes  the dominant hadronic channel.

{\bf For  2HDM-Y} the  ratio of hadronic
widths depends on $\tan\beta$. At $\tan\beta<6$, the term
$F_U^2$ dominates as in previous case, and ratio of widths is
given by eq.~\eqref{RatioIX},  Fig.~\ref{FigratIX}-A. With the growth of $\tan\beta$ relative role of $F_D$ term increases, at
$\tan\beta>8$ the term $F_D^2$ becomes dominant and we have (see Fig.~\ref{FigratIX}-B)
\bear{c} \fr{\Gamma(H^+\to\bar{b}t^*\to
\bar{b}bW^+)}{\Gamma(H^+\to
cs)}%=\\[3mm]
=\left(\fr{M_b}{M_s}\right)^2\fr{T\alpha}{8\pi
s_W^2}\mathlarger\int\limits_1^y
\fr{(y-x)^2(x-1)^2(2+x)}{y^2x^2(T-x)^2}\,dx.
\eear{RatioY}
The ratio of main couplings is reduced from 135 at $\tan\beta<6$ to $ M_b/M_s\sim 40$ at $\tan\beta\ge 9$.                                                              Due to this reduction, the  area, where 3-body channel is essential, narrows. Note that the narrowing of the range of dominance of 3-body decay occurs in relatively narrow interval of  $6<\tan\beta<9$. In this interval contribution of the interference term $F_UF_D$ is also essential.

The comparison with leptonic decay channel ($H^+\to \tau^+\nu$) can be made in the same  manner.

{\bf Summary}. At $M_+<135$~GeV, only 2-body decays of $H^+$ can be observed.
At 180~GeV$>M_+>135$~GeV, one should try to observe 3-body decay as well.
The BR's for different decay channels ($cs$, $\tau\nu$, $Wbb$) depend  on the model of Yukawa sector and value of $tan\beta$. Therefore, even rough measuring of BR's for these decay channels will give important information about mentioned properties of the model. \\

I am  thankful A. Djouadi, E. Ma,
V.G.~Serbo and M.I.~Vysotsky for discussions and V.S. Cherkassky for help in plotting. This paper is supported by grants
RFBR 11-02-00242-à, NSh-3802.2012.2, Program of Dept. of Phys. Sc
RAS and SB RAS "Studies of Higgs boson and exotic particles at
LHC" and  by Polish Ministry of Science and Higher Education Grant
 N202 230337.

\end{document}